\documentclass[fleqn,usenatbib]{mnras}

\usepackage{newtxtext,newtxmath}

\usepackage[T1]{fontenc}

\DeclareRobustCommand{\VAN}[3]{#2}
\let\VANthebibliography\thebibliography
\def\thebibliography{\DeclareRobustCommand{\VAN}[3]{##3}\VANthebibliography}

\usepackage{graphicx}	
\usepackage{amsmath}	

\usepackage{array}      

\newcommand{\nus}{{\em NuSTAR}}
\newcommand{\suz}{{\em Suzaku}}
\newcommand{\ixpe}{{\em IXPE}}
\newcommand{\rc}{R_{\textrm{c}}}
\newcommand{\rg}{$R_{\textrm{G}}$}
\newcommand{\rin}{R_{\textrm{in}}}
\newcommand{\monk}{\textsc{monk}}
\newcommand{\kte}{kT_{\textrm{e}}}
\newcommand{\expo}[2]{$ #1 \times 10^{#2}$}
\newcommand{\ser}[2]{$#1 \pm #2$}
\newcommand{\fluxcgs}{erg~cm$^{-2}$~s$^{-1}$}
\newcommand{\sigmat}{\sigma_{\textrm{T}}}
\newcommand{\compps}{{\sc compps}}

\title[Differentiating between coronal geometries in AGNs with \ixpe]{Prospects for differentiating extended coronal geometries in AGNs with the \ixpe\ mission}

\author[F. Ursini et al.]{
F. Ursini,$^{1}$\thanks{E-mail: francesco.ursini@uniroma3.it}
G. Matt,$^{1}$
S. Bianchi,$^{1}$
A. Marinucci,$^{2}$
M. Dov\v{c}iak,$^{3}$
and
W. Zhang$^{4}$
\\
$^{1}$	Dipartimento di Matematica e Fisica, Universit\`a degli Studi Roma Tre, via della Vasca Navale 84, 00146 Roma, Italy\\
$^{2}$ ASI - Italian Space Agency, Via del Politecnico snc, 00133, Rome, Italy\\
$^{3}$Astronomical Institute, Academy of Sciences of the Czech Republic, Bo\v{c}n\'i II 1401, CZ-14100 Prague, Czech Republic\\
$^{4}$National Astronomical Observatories, Chinese Academy of Sciences, 20A Datun Road, Beijing 100101, China
}

\date{Accepted XXX. Received YYY; in original form ZZZ}

\pubyear{2021}

\begin{document}
\label{firstpage}
\pagerange{\pageref{firstpage}--\pageref{lastpage}}
\maketitle

\begin{abstract}
X-ray polarimetry can potentially constrain the unknown geometrical shape of AGN coronae. We present simulations of the X-ray polarization signal expected from AGN coronae, assuming three different geometries, namely slab, spherical and conical. We use the fully relativistic Monte-Carlo Comptonization code \monk\ to compute the X-ray polarization degree and angle. We explore different coronal parameters such as shape, size, location and optical depth. Different coronal geometries give a significantly different X-ray polarization signal. A slab corona yields a high polarization degree, up to 14\% depending on the viewing inclination; a spherical corona yields low values, about 1--3\%, while a conical corona yields intermediate values. We also find a difference of 90 degrees in polarization angle between the slab corona and the spherical or conical coronae. Upcoming X-ray polarimetry missions like \ixpe\ will allow us to observationally distinguish among different coronal geometries in AGNs for the first time.
\end{abstract}



\section{Introduction}
\label{sec:intro}
An X-ray corona is one of the main constituents of active galactic nuclei (AGNs) and primary contributor to their X-ray emission \cite[e.g.][]{haardt&maraschi1991}. The corona is situated in the innermost region of the accretion flow and in the close proximity to the event horizon of the central supermassive black hole \cite[e.g.]{reis&miller2013}. Indeed, microlensing studies show that the size of the X-ray corona is of the order of a few gravitational radii \cite[e.g.][]{chartas2009,chartas2016}. The X-ray emission of AGNs thus provides a powerful probe of General Relativity in the strong gravity regime. 
However, the physical origin of the corona is a matter of debate, although magnetic processes are often invoked for its formation and heating \cite[e.g.][]{dimatteo1998}.
The origin of the corona is related to its geometry, which is essentially unknown. The corona could be a sphere-like region above the black hole, possibly powered by magnetic reconnection \citep[e.g.][]{wilkins&fabian2012}, or it could have a conical shape if it forms the base of a jet \cite[e.g.][]{henri1991,hp1997,markoff2005} or a failed jet \citep{ghm2004}. The corona could also be a slab-like structure sandwiching the accretion disc \cite[e.g.][]{haardt&maraschi1993}, perhaps originating from magnetic instabilities \cite[e.g.][]{dimatteo1998}.
Constraining the coronal geometry is thus of prime importance to shed light on the physics of the disc-corona system. 

Currently, constraints on the coronal geometry can only be derived from spectral and/or timing properties, such as the analysis of time lags between different energy bands \cite[e.g.][]{demarco2013,kara2016,caballero2020}. 
In principle, polarimetry could yield more direct and model-independent measurements. Indeed, the polarization properties of the radiation crucially depend on the geometry of the emitting system. In AGNs, the primary emission from the corona could be significantly polarized \citep[e.g.][]{haardt&matt1993,pv1993,SK2010,dovciak2011,tamborra2018}. 
The polarization of X-rays reprocessed by the disc and surrounding material can be a powerful probe of the disc/corona geometry \citep[e.g.][]{matt1989,matt1993}, especially in Seyfert 2 galaxies, in which relatively high polarization degrees are expected \citep{marin2018_sy2}. Seyfert 1 galaxies, on the other hand, offer a more direct view of the corona: their 2--10 keV emission is usually dominated by the primary power law, and its polarimetric properties can be used to constrain the geometrical shape of corona \citep[e.g.][]{beh2017}.

X-ray polarimetric studies of AGNs will become possible for the first time thanks to the Imaging X-ray Polarimetry Explorer \cite[\ixpe,][]{ixpe}, a NASA/ASI mission that has been successfully launched on December 9, 2021. \ixpe\ is the first X-ray mission dedicated to polarimetry, carrying three X-ray telescopes with polarization-sensitive imaging detectors \cite[the gas-pixel detectors,][]{costa2001} operative in the 2--8 keV band \citep{ixpe}.
\ixpe\ is expected to perform meaningful measurements of the X-ray polarization of different types of sources, including AGNs \citep{ixpe,mw2017}. 
An X-ray polarimetry array will also be a key element of the enhanced X-ray Timing and Polarimetry mission \cite[eXTP,][]{extp}, which is planned for launch in 2027. 

In this work, we present numerical simulations of the X-ray polarimetric signal expected from the corona of radio-quiet, unobscured type 1 AGNs, using the general relativistic Monte Carlo radiative transfer code \monk\ (\citealt{zhang2019}; Zhang et al., submitted). 
An analogous approach has been followed by \cite{beh2017}, who simulated the broad-band X-ray polarization signal of Seyfert 1 galaxies, for two corona geometries (wedge and spherical shell) and three different corona sizes. 
Our aim is to determine whether different coronal geometries can be distinguished with the polarimetric analysis of the primary X-ray emission, with an emphasis on upcoming \ixpe\ observations.  
The paper is structured as follows. We discuss the application of the \monk\ code and the numerical setup in Sect. \ref{sec:setup}. We present the results in Sect. \ref{sec:results} and summarize our conclusions in Sect. \ref{sec:conclusions}.

\section{Setup}\label{sec:setup}
\monk\ calculates the energy and polarization spectra of Comptonized radiation from a corona illuminated by a standard accretion disc \citep{NT}. Here we briefly summarize the procedure implemented in \monk, referring to \cite{zhang2019} for a detailed description. First, optical-UV seed photons are generated according to the disc emissivity, with initial polarization given by the calculation of \cite{chandra1960} for a semi-infinite planar atmosphere. Then, the photons are ray-traced along null geodesics in Kerr spacetime, propagating the polarization vector. Photons reaching the corona are Compton scattered assuming the Klein-Nishina cross section. The Stokes parameters of the scattered photons are computed in the electron rest frame \citep{connors1980} and then transformed in the observer (Boyer-Lindquist) frame. The propagation terminates when the photons either enter the event horizon, hit the disc or arrive at infinity. Counting the latter photons, the energy and polarization spectrum is constructed. 
Since scattering produces linearly polarized photons, the code computes the Stokes parameters $Q$ and $U$, while $V$ is set at zero.

The main input parameters of \monk\ are: the black hole mass and spin; the accretion rate; the physical (optical depth and temperature) and geometrical parameters of the corona. 
In our simulations, we set a black hole mass of \expo{2}{7} solar masses and an Eddington ratio of 0.1. These parameters are chosen to be consistent with
MCG-5-23-16, an AGN optically classified as a Seyfert 1.9 \citep{veron1980} with broad emission lines in the infrared \citep{vv2010,onori1}, likely viewed at an inclination of $\sim 50$ deg \citep[e.g.][]{zoghbi2017}. This source is an excellent target for \ixpe, being X-ray bright and Compton-thin \citep{balokovic2015,zoghbi2017}.
Its black hole mass is estimated from the X-ray variability \citep{caixa3} and is consistent with the virial mass estimated from the infrared lines \citep{onori2}. The Eddington ratio is set to 0.1 in agreement with the observed luminosity \citep{zoghbi2017}.
Concerning the black hole spin $a$, we make simulations assuming the two possible values $a=0$ and 0.998. 

\begin{figure}
	\centering
	\includegraphics[width=0.66\columnwidth]{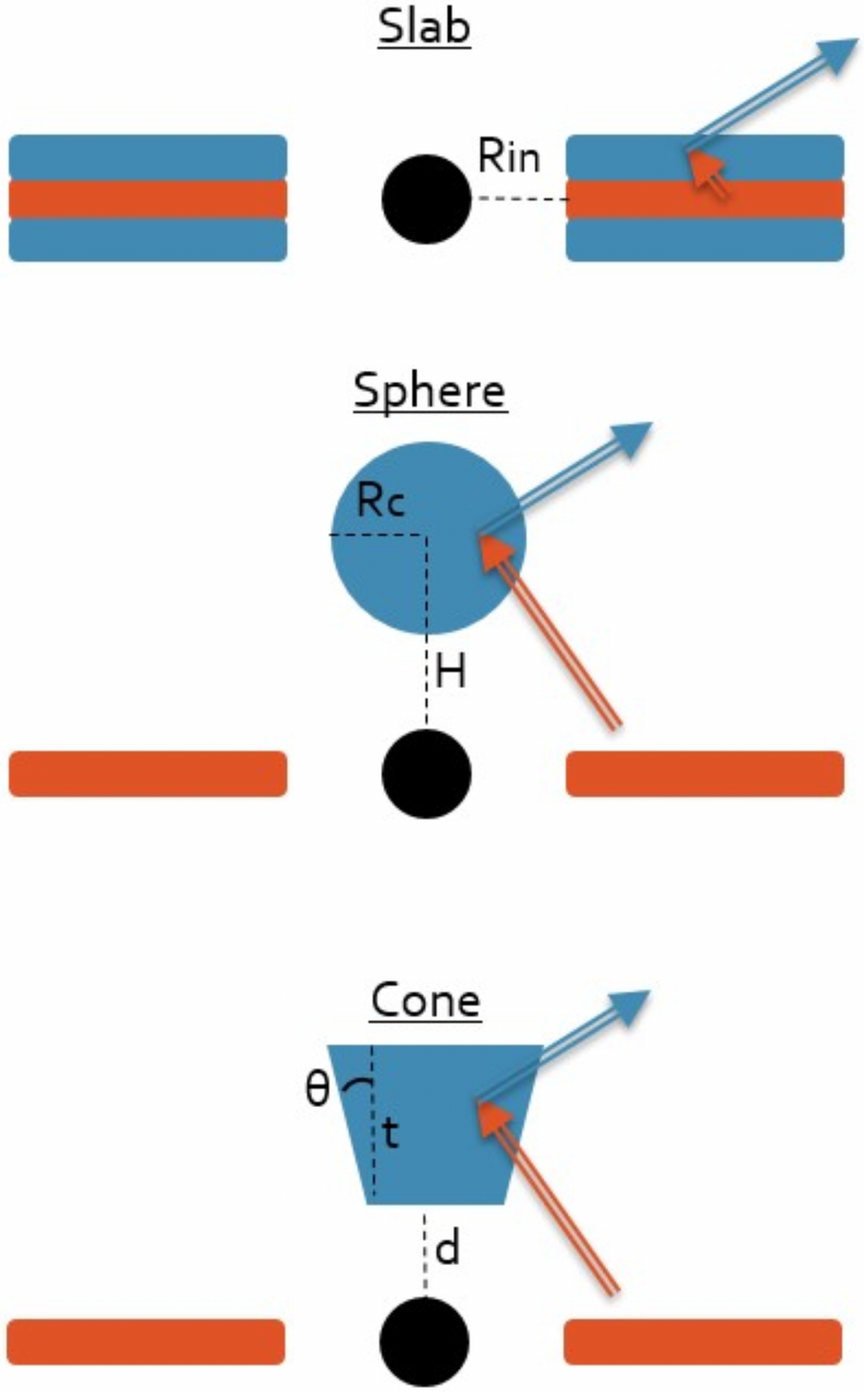}
	\caption{Different coronal geometries. The X-ray emission from the corona (in light blue) is produced by inverse Compton scattering of the optical--UV radiation from the disc (in orange). \label{fig:coronae}}
\end{figure}

Different coronal geometries can be assumed in \monk\ \citep{zhang2019}. Here we focus on three alternative configurations, namely the slab, the spherical lamppost, and the truncated cone, as we describe in the next sections (Fig. \ref{fig:coronae}). 
Concerning the physical parameters of the corona, \cite{balokovic2015} reported measurements of the coronal optical depth $\tau$ and temperature $\kte$ based on \nus\ data of MCG-5-23-16. \cite{zoghbi2017} also reported photon indices of 1.8--1.9 from the analysis of \suz\ and \nus\ data of MCG-5-23-16 spanning 10 years. 
The \nus\ spectra exhibit a well constrained high-energy cut-off between 100 and 200 keV \citep{balokovic2015,zoghbi2017}.
However, to obtain more general results and explore the relation between the polarimetric properties and the spectral shape, we proceed as follows. We assume three possible values of the photon index $\Gamma$ of the primary power law in the 2--8 keV band, namely 1.6, 1.8, and 2. Then, for each value of $\Gamma$, we set three pairs $\tau,\kte$ consistent with it. In other words, we set $\tau,\kte$ such that the Comptonization spectrum produced by \monk\ is (a posteriori) consistent with a power law having the chosen photon index.

\subsection{Slab}
The slab corona is assumed to fully cover the disc, that can either be extended down to the innermost stable circular orbit (ISCO) or truncated at an arbitrary radius. In general, the ISCO depends on the black hole spin \citep[e.g.][]{mtw}. We assume three possible values of the inner disc radius $\rin$ (see also Table \ref{tab:geom}): (i) $a=0.998$, $\rin =$ ISCO $\equiv 1.24$ in units of gravitational radii (\rg = $GM/c^2$); (ii) $a=0$, $\rin =$ ISCO $\equiv 6$ \rg; (iii) $a=0$, $\rin=30$ \rg\ (truncated case). The vertical thickness of the slab corona is also a parameter of \monk, and we set it to 1 \rg\ in all cases \cite[see also][]{zhang2019}. However, the results are not strongly dependent on this parameter. The physical parameters are summarized in Table \ref{tab:phys}: we set $\kte=25, 50$ or 100 keV and an optical depth $\tau$ consistent with the assumed photon index. In \monk, the optical depth in slab geometry is defined as $\tau=n_e \sigmat h$ where $h$ is the half-thickness of the slab\footnote{This definition is the same as in the popular Comptonization codes \textsc{comptt} \citep{titarchuk1994} and \compps\ with $\mathtt{cov\_frac} \neq 1$ \citep{compps}.} \citep{zhang2019}. The values of $\tau$ in our simulations range between 0.25 and 1.65 (see Table \ref{tab:phys}). In the slab case, we assume the corona to be co-rotating with the Keplerian disc \citep{zhang2019}.

\subsection{Spherical lamppost}
A spherical lamppost corona is characterized by the height $H$ above the
disc and the coronal radius $\rc$.
For the \monk\ simulations, we choose a few values of these two parameters following the estimates of the corona size by \cite{csize}, obtained using the relativistic ray-tracing code of \cite{dovciak&done2016}. These estimates are based on a simple argument: the corona must intercept a photon flux from the disc that is consistent with the observed X-ray flux, because Comptonization conserves the number of photons. The estimates of the lamppost corona size depend also on the spin; indeed, a larger spin allows for more compact corona closer to the event horizon \citep{csize}. We choose the following values of the coronal height and radius (see also Table \ref{tab:geom}): (i) $a=0.998$, $H=5$ \rg, $\rc = 2$ \rg; (ii) $a=0$, $H=10$ \rg, $\rc = 7$ \rg; (iii) $a=0$, $H=30$ \rg, $\rc = 10$ \rg. The optical depth in spherical geometry is the radial one, and its values in our simulations range between 0.8 and 4.5 (see Table \ref{tab:phys}). Finally, in the spherical case we assume a stationary corona \citep{zhang2019}.

\subsection{Truncated cone}
If the corona is outflowing rather than static, for example forming the base of a jet, it might be better described as a conical blob opening away from the black hole.
Since we focus on radio-quiet (i.e. non-jetted) Seyferts, following \cite{ghm2004} we assume that the corona is a failed jet with a sub-relativistic bulk velocity, smaller than the escape velocity. In the scenario proposed by \cite{ghm2004}, blobs of material are launched with some initial velocity and then fall back, actually failing to produce a jet akin to that of radio-loud AGNs. In \monk, this type of corona is schematized as a single blob with a truncated conical shape, located on the symmetry axis of the disc, formed by electrons with a bulk velocity $\beta$ (see Fig. \ref{fig:coronae}, bottom panel).
For consistency with the failed jet scenario, in our simulations we assume $\beta=0.3$, which is less than the escape velocity at initial radii of 20-30 \rg\ \citep{ghm2004}.
The three relevant geometrical parameters are the distance $d$ between the lower base and the disc, the intrinsic height or thickness $t$ of the truncated cone, and its half-opening angle $\theta$. The vertex of the cone is at the center of the black hole, so that the radius of the lower base is $d \tan \theta$.
The optical depth in \monk\ is proportional to the diameter of the lower base, being defined as $\tau=n_e \sigma_T 2 d \tan \theta$.  
Concerning the half-opening angle, we assume $\theta=30$ deg.
The results are not much altered using different parameters, such as $\beta=0.1$ or $\theta=60$ deg, as long as $\beta$ is not too close to unity.

\begin{table}
	\centering
	\caption{Geometrical parameters of the corona in the \monk\ simulations (in units of \rg) for different values of the black hole spin $a$.\label{tab:geom}}
	\begin{tabular}{l |c | c c | c c}
		\hline \hline
		$a$ & \multicolumn{1}{c|}{slab}& \multicolumn{2}{c|}{sphere} & \multicolumn{2}{c}{cone}\\
		\hline
		&$\rin$&	$H$ & $\rc$&$d$ & $t$\\
		\hline
		0.998 & 1.24 & 5 & 2 & 3 & 10\\
		0 & 6&10&7&5&15\\
		0 & 30&30&10&20&20\\
		\hline
	\end{tabular}
\end{table}

\begin{table}
	\centering
	\caption{Physical parameters of the corona. The optical depth for the slab is vertical and measured from the half-plane of the disc, while it is radial for the spherical lamppost, and it depends on the diameter of the upper base for the truncated cone. The coronal temperature $\kte$ is in keV. \label{tab:phys}}
	\begin{tabular}{l |c c | c c | cc}
		\hline \hline
		$\Gamma$&\multicolumn{2}{c|}{slab}& 	\multicolumn{2}{c|}{sphere} & \multicolumn{2}{c}{cone}\\
		\hline
		&$\tau$ & $\kte$ & $\tau$ & $\kte$ & $\tau$ & $\kte$  \\
		\hline
		&0.55&100&1.5&100&1&100\\
		1.6&1&50&2.7&50&2&50\\
		&1.65&25&4.5&25&3.2&25\\
		\hline
		&0.35&100&1&100&0.7&100\\
		1.8&0.75&50&2&50&1.3&50\\
		&1.35&25&3.5&25&2.3&25\\
		\hline
		&0.25&100&0.8&100&0.5&100\\
		2&0.6&50&1.6&50&1&50\\
		&1.15&25&3&25&1.7&25\\
		\hline
	\end{tabular}
\end{table}

\section{Results}\label{sec:results}
We run 81 simulations with the parameters described above. To compute the polarization properties, for each simulation we assume an observer at different inclination angles, between 0 and 90 degrees \cite[for a discussion of the dependency of the spectrum on the inclination, see][]{zhang2019}. We integrate the signal in the \ixpe\ bandpass, namely 2--8 keV.

In the following sections,
we plot the polarization degree and angle as a function of the cosine of the inclination angle $\mu$. The polarization angle is measured from the north-south direction in the sky plane, meaning that a polarization vector parallel to the disc corresponds to a polarization angle of 90 degrees.
Below we focus on those coronal parameters that are in better agreement with the X-ray properties of MCG-5-23-16. We present in the appendix \ref{appendix} the complete results, for all the assumed geometrical parameters and the different black hole spins.

\subsection{Polarization degree}
\begin{figure*}
	\includegraphics[width=\textwidth]{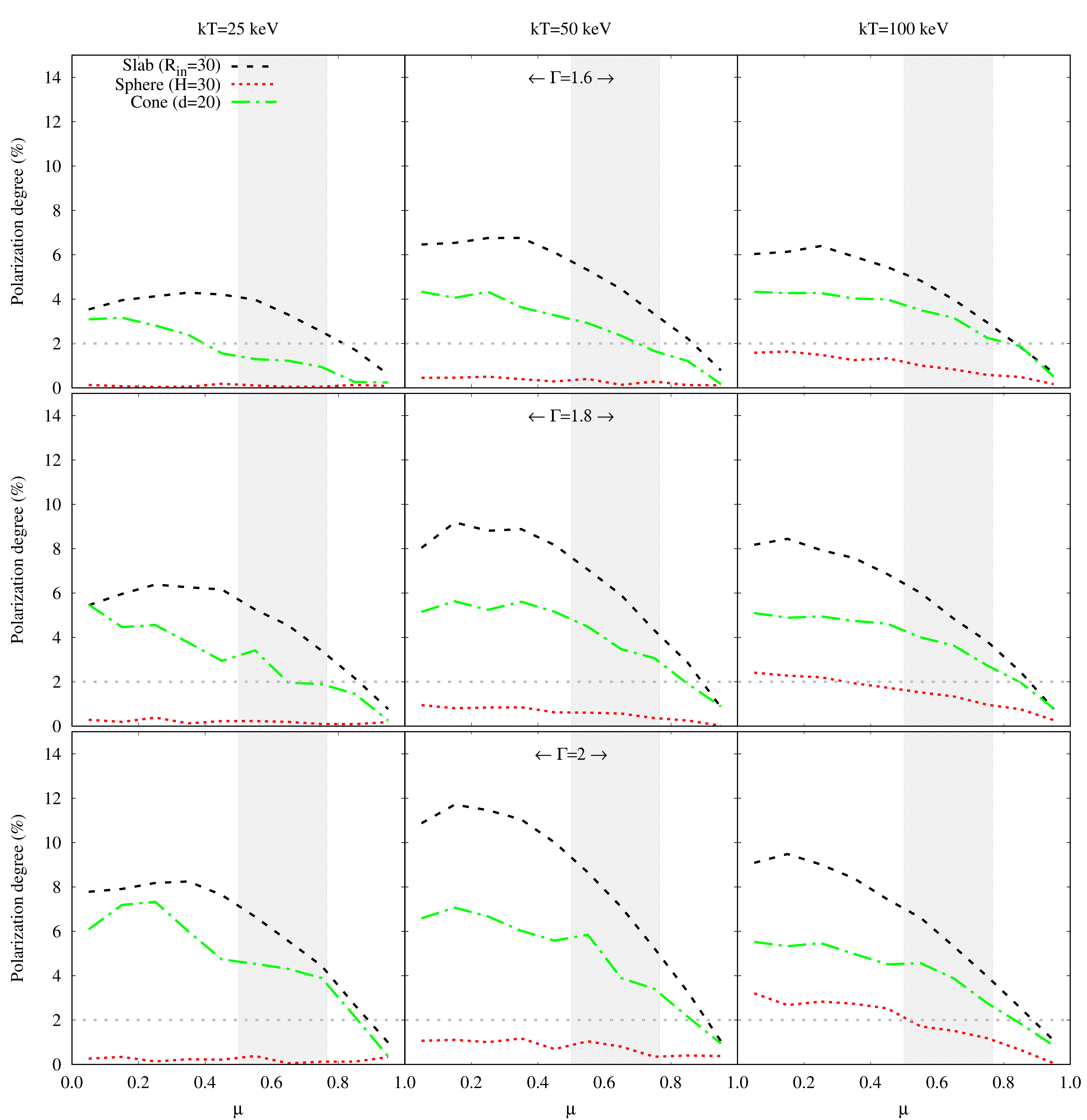}
	\caption{
			Comparison between the polarization degrees (2--8 keV) for the slab, spherical, and conical geometry, plotted versus the cosine of the inclination angle of the observer. The different rows show the results for the different X-ray photon indices assumed (from top to bottom: $\Gamma=1.6,1.8,$ and 2), while the different columns correspond to the different coronal temperatures (from left to right: $\kte=25, 50$ and 100 keV). The shaded area corresponds to the inclination interval 40--60 deg, while the gray dotted line shows the minimum polarization detectable by \ixpe\ in a 500 ks exposure of MCG-5-23-16.
		\label{fig:slab-sphere}}
\end{figure*}
In Fig. \ref{fig:slab-sphere}, we show a comparison between the polarization degree for the slab, the sphere, and the cone.
We plot the results for the truncated slab geometry, which is probably more consistent with the X-ray spectrum of MCG-5-23-16 \citep[based on the reflection component, see][]{zoghbi2017}. For the sphere, we plot the case with $H=30$ \rg, which is anyway similar to the others (Fig. \ref{fig:deg-sphere}, central row). For the cone, we plot the case with $d=20$ \rg\ (Fig. \ref{fig:deg-cone}, center right panel).
Although the inclination of the source is uncertain, a value of \ser{50}{10} deg is consistent with the properties of the X-ray reflection component \citep{weaver1998,braito2007,reeves2007,guainazzi2011,zoghbi2017}. For this inclination interval,
and for the case $\Gamma = 1.8$,
the polarization degree is $\sim 4{-}8\%$ for the slab, $\sim 1\%$ for the sphere, and around 2--5\% for the cone (Fig. \ref{fig:slab-sphere}, central row).
In any case, the highest polarization degrees are obtained for the slab corona, with values up to 12\%, while for a spherical corona the polarization degree is always below 3\%. The conical corona produces intermediate values, in any case below 8\%.

\begin{figure*}
	\includegraphics[width=\textwidth]{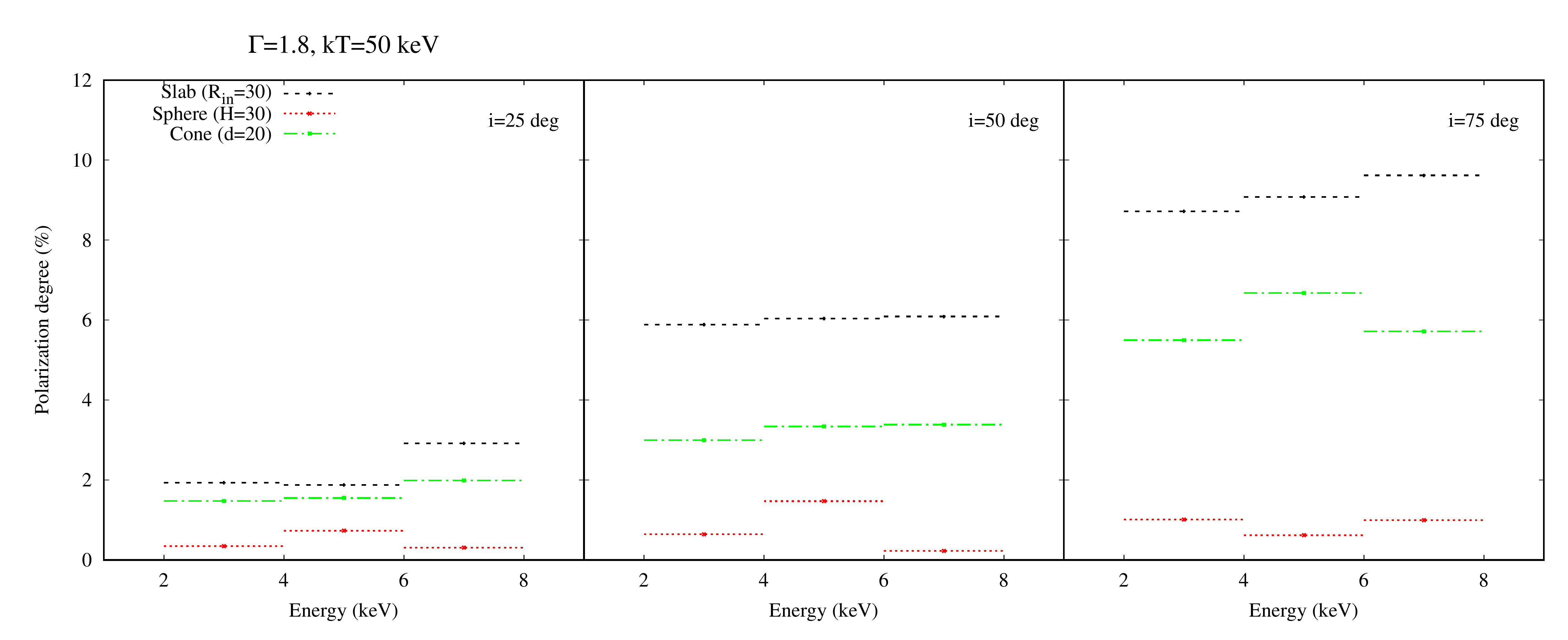}
	\caption{Polarization degree versus energy, for three different geometries in the case $\Gamma=1.8$, $\kte=50$ keV. Each plot correspond to a different value of the inclination (from left to right: 25, 50 and 75 deg). 
		\label{fig:polspec}}
\end{figure*}

Using the same geometries as above, we plot in Fig. \ref{fig:polspec} the polarization degree as a function of the energy, in the 2--8 keV band, now divided into three energy bins. 
For simplicity, we plot the results only for $\mathbf{\Gamma=1.8}$, $\kte=50$ keV and for three different values of the inclination, namely 25, 50 and 75 deg, respectively.
In all cases, the slab corona produces the largest polarization degrees.

\subsection{Polarization angle}\label{subsec:angle}
\begin{figure*}
	\includegraphics[width=\textwidth]{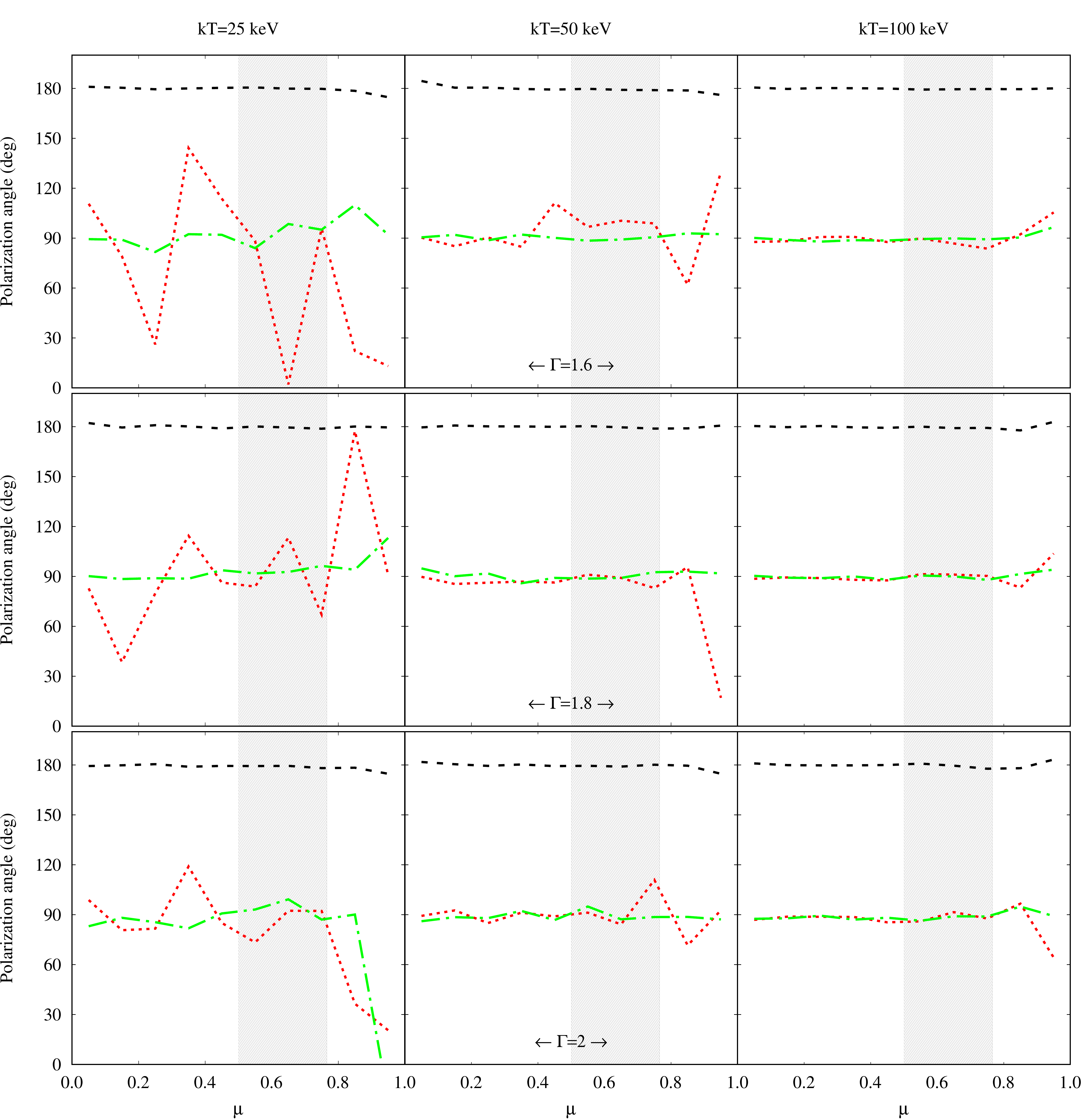}
	\caption{
			Polarization angle (in degrees) for the slab, spherical, and conical geometry, versus the cosine of the inclination angle. Line and color coding are the same as Fig. \ref{fig:slab-sphere}.
		\label{fig:slab-sphere-ang}}
\end{figure*}

The polarization angle is also a crucial observable parameter. For the slab, it is always close to 180 deg, while it is mostly scattered around 90 deg for the spherical and conical geometries (Fig. \ref{fig:slab-sphere-ang}). In these two latter cases, 
when the polarization degree is very low, the scatter in polarization angle can be quite large. This indicates a large uncertainty in the polarization angle when the observed radiation is almost unpolarized.

\begin{figure*}
	\includegraphics[width=\textwidth]{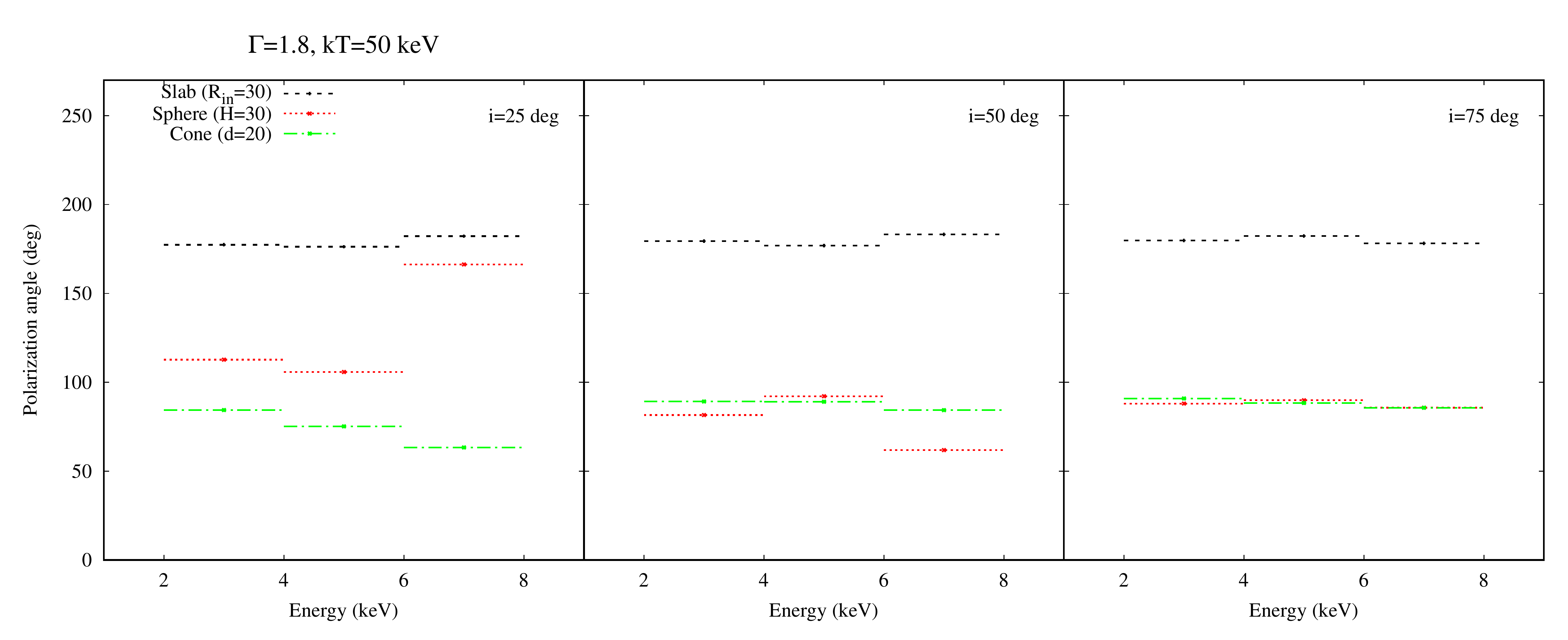}
	\caption{Polarization angle versus energy, for the same geometries as in Fig. \ref{fig:polspec}. 
		\label{fig:angspec}}
\end{figure*}

In Fig. \ref{fig:angspec}, we plot the polarization angle as a function of the energy, for three different geometries, like in Fig. \ref{fig:polspec}. We note that the difference in polarization angle between the slab and the spherical/conical corona is not strongly dependent on the energy, nor on the inclination.

\section{Discussion and conclusions}\label{sec:conclusions}
The geometrical shape of the X-ray corona of AGNs cannot be constrained via spectroscopy. Thanks to future missions like \ixpe, X-ray polarimetry will open a new observational window, adding two observable parameters: the polarization degree and the polarization angle.  
We performed simulations with the relativistic Monte-Carlo Comptonization code \monk, showing that the polarimetric signal expected from AGN coronae is significantly different depending on the coronal geometry.
We focused on the case of MCG-5-23-16, a very promising candidate for upcoming \ixpe\ observations. 
Our main results are summarized in Table \ref{tab:res} (see also Appendix \ref{appendix}).

A slab corona yields relatively large polarization degrees, up to 12\% depending on the inclination. A spherical corona mostly yields very low polarization degrees, below 1\%; we obtain values of 2-3\% only for the most extended (radius of 10 \rg) and distant (height of 30 \rg\ above the disc) case, 
as shown in Fig. \ref{fig:slab-sphere}.
The reason could be that a more extended/distant lamppost corona is more illuminated from the bottom, meaning that the seed photon distribution is less isotropic. However, even in this case, the polarization degree is less than in the slab configuration. A conical corona, on the other hand, yields polarization degrees (for a given inclination) mostly between the slab and spherical cases. 
For a given set of physical parameters, the polarization degree also depends on the size of the corona \citep[see also][]{beh2017,tamborra2018} which in turn depends on the black hole spin \citep{dovciak&done2016,csize}.
This is more clearly shown in the Appendix \ref{appendix}.

\begin{table}
	\centering
	\caption{Summary of the results (polarization degree and angle) for the three coronal geometries. \label{tab:res}}
	\begin{tabular}{l c c}
		\hline \hline
		&pol. degree& pol. angle\\
		\hline
		slab&high (up to 12\%)&$\sim180$ deg\\
		sphere&low (1--3\%)&$\sim 90$ deg\\
		cone&intermediate (up to 7\%)&$\sim90$ deg\\
		\hline
	\end{tabular}
\end{table}

According to these results, distinguishing between the different geometries is well within the capabilities of \ixpe\ \citep[see also][]{marinucci2019}. In general, the sensitivity of a polarimeter is quantified by the minimum detectable polarization (MDP). The MDP at 99\% confidence level is \citep[e.g.][]{weisskopf2010}:
\begin{equation}
\textrm{MDP}_{99} = \frac{4.29}{M \, S} \sqrt{\frac{S + B}{T}}
\end{equation}
where $M$ is the modulation factor, $S$ is the source count rate, $B$ is the background count rate and $T$ is the observation length. For \ixpe, the MDP$_{99}$ is 2\% for an observation of 500 ks of an AGN with a 2--10 keV flux of \expo{1}{-10} \fluxcgs\ such as MCG-5-23-16 (assuming $\Gamma=1.8$). We can also estimate the uncertainty on the measurement of the polarization degree, following \cite{kislat2015}:
\begin{equation}
\sigma_p \simeq \sqrt{\frac{2}{S M^2}}.
\end{equation}
The same \ixpe\ observation as above would yield an uncertainty of about 0.7\%. 

In combination with the polarization degree, the polarization angle may allow us to break the degeneracy between different geometries with high significance, if the orientation of the system is known or at least can be assumed from independent measurements. Indeed, in most cases there is a difference of 90 degrees among the slab and the spherical or conical coronae. In the latter two cases, the polarization angle is not well constrained when the polarization degree is very low. This happens especially for high values of the coronal optical depth \cite[see also][]{tamborra2018}. However, even in this case, the slab geometry can be ruled out.

It is important to remark that the primary X-ray emission
from the corona can be reprocessed by the accretion disc
or the dusty torus at larger distances. This produces a reflection component, ubiquitously observed in the X-ray spectrum of Seyferts, consisting of a fluorescence iron line and a Compton hump peaking at 20-30 keV  \cite[e.g.][]{george&fabian1991,matt1991}.  
Theoretical models predict a degree of polarization  of the reflected component as high as $\sim 30 \%$ \citep{matt1989,dovciak2004,marin2018_sy2}. 
However, the actual contribution to the total polarization degree depends on the relative strength of reflection compared with the primary emission, which is expected to be low in the energy range typical of photoelectric polarimeters. 
Given the values measured in Compton-thin AGNs \citep[e.g.][]{zappacosta2018,panagiotou2019}, 
the contribution of the Compton reflection component to the total 2--8 keV flux is 5-10\%.
This means that the contribution to the polarization degree in this band should be no more than 3\%, at least for Compton-thin AGNs \cite[see also][]{marin2018_sy1}. Finally, the net observed polarization critically depends on the relative orientation of the polarization pseudovectors, meaning that the polarization degree increases if the components are parallel and decreases if they are orthogonal. 
Despite these complexities, broad-band spectroscopy allows us to properly disentangle the reflection component from the primary continuum, as in the case of MCG-5-23-16 \citep{zoghbi2017}. This will in turn allow us to assess the different contributions to the polarization spectrum with empirical fits.
This approach is beyond the scope of this paper, however the simulations discussed here will be a key ingredient  to robustly model the X-ray polarimetric signal of AGNs. 

We note that measuring the physical parameters of the corona is of prime importance to restrict the parameter space and properly compare theoretical predictions with observations. A bright source like MCG-5-23-16, which is seen at an intermediate inclination and has a quite standard X-ray photon index, is optimal to constrain the polarization signal and, in so doing, the coronal geometry. This will in turn constrain the physical origin of the corona, which is still an open problem.

\section*{Acknowledgements}
We thank the referee for useful suggestions that significantly improved the quality of the manuscript.
FU, SB and GM acknowledge financial support from the Italian Space Agency
(grants 2017-12-H.0 and 2017-14-H.0).
	WZ acknowledges the support by the Strategic Pioneer Program on Space Science, Chinese Academy of Sciences through grant XDA15052100.

\section*{Data Availability}

The code underlying this article, \monk, is proprietary. Simulation data supporting the findings of the article will be shared on reasonable request.


\bibliographystyle{mnras}
\bibliography{mybib} 


\appendix

\section{Complete results}\label{appendix}
\begin{figure*}
	\centering Slab geometry \par\smallskip
	\includegraphics[width=\textwidth]{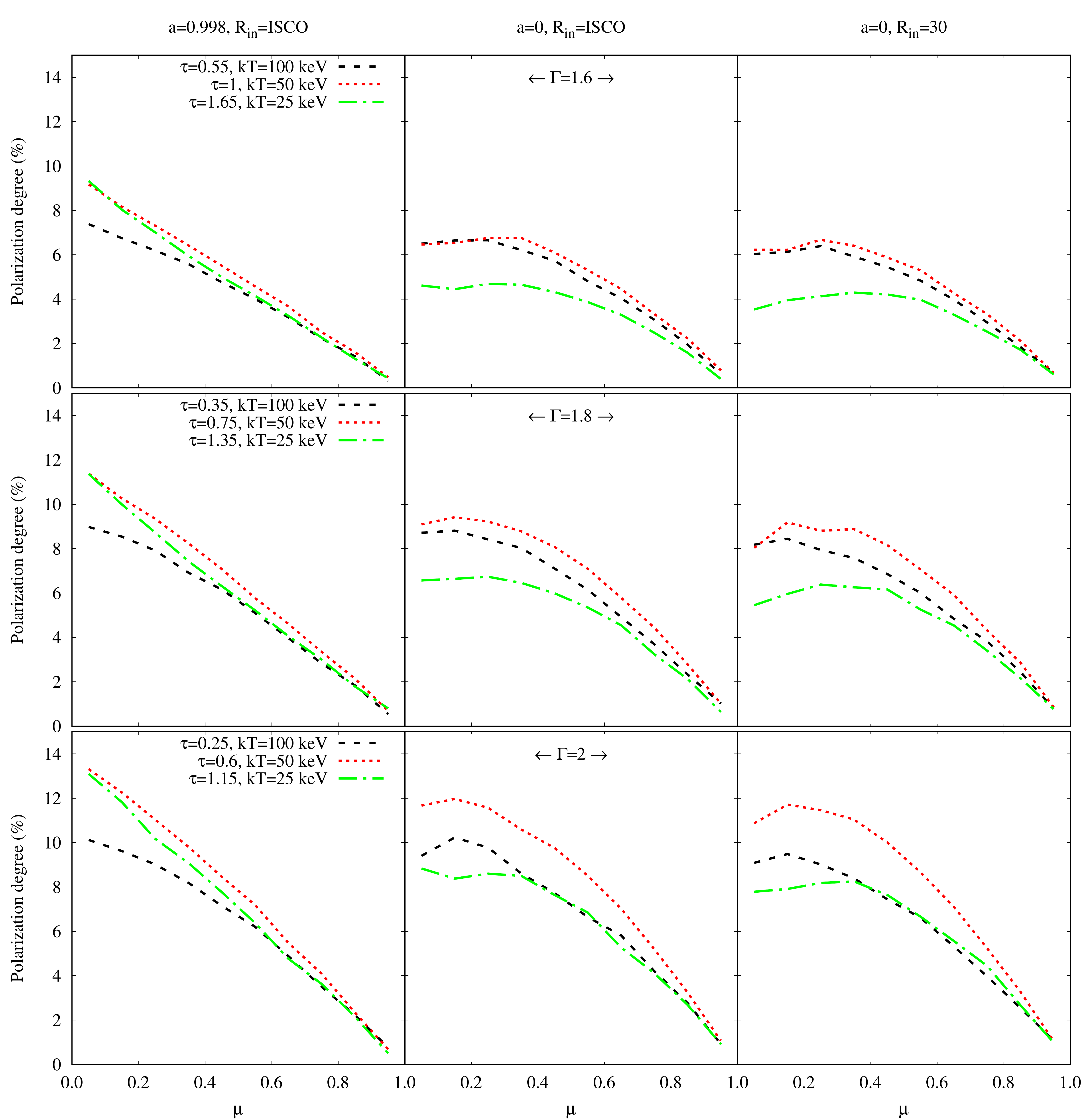}
	\caption{Polarization degree in the 2--8 keV band versus the cosine of the inclination angle of the observer, for the slab geometry. The different rows show the results for the different photon indices assumed (from top to bottom: $\Gamma=1.6,1.8,$ and 2); colors and dash types highlight the different pairs $\tau,\kte$ reproducing the photon index (see also Table \ref{tab:phys}). Column-wise, the plot shows the results for the different black hole spin and/or inner radius of the slab (see also Table \ref{tab:geom}. \label{fig:deg-slab}}
\end{figure*}

\begin{figure*}
	\centering Slab geometry \par\smallskip
	\includegraphics[width=\textwidth]{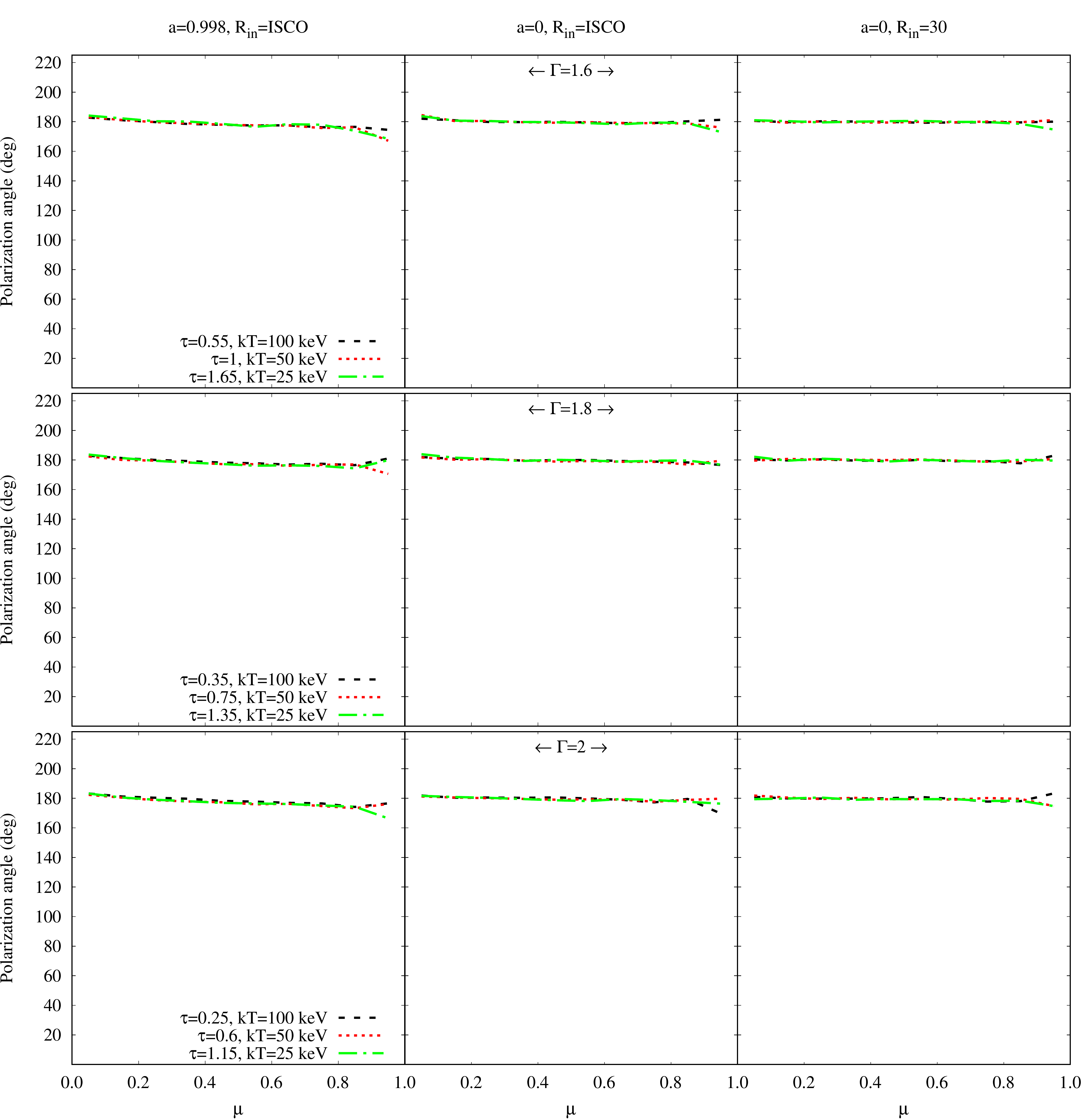}
	\caption{Polarization angle versus the cosine of the inclination angle, for the slab geometry. \label{fig:ang-slab}}
\end{figure*}

\begin{figure*}
	\centering Spherical lamppost geometry \par\smallskip
	\includegraphics[width=\textwidth]{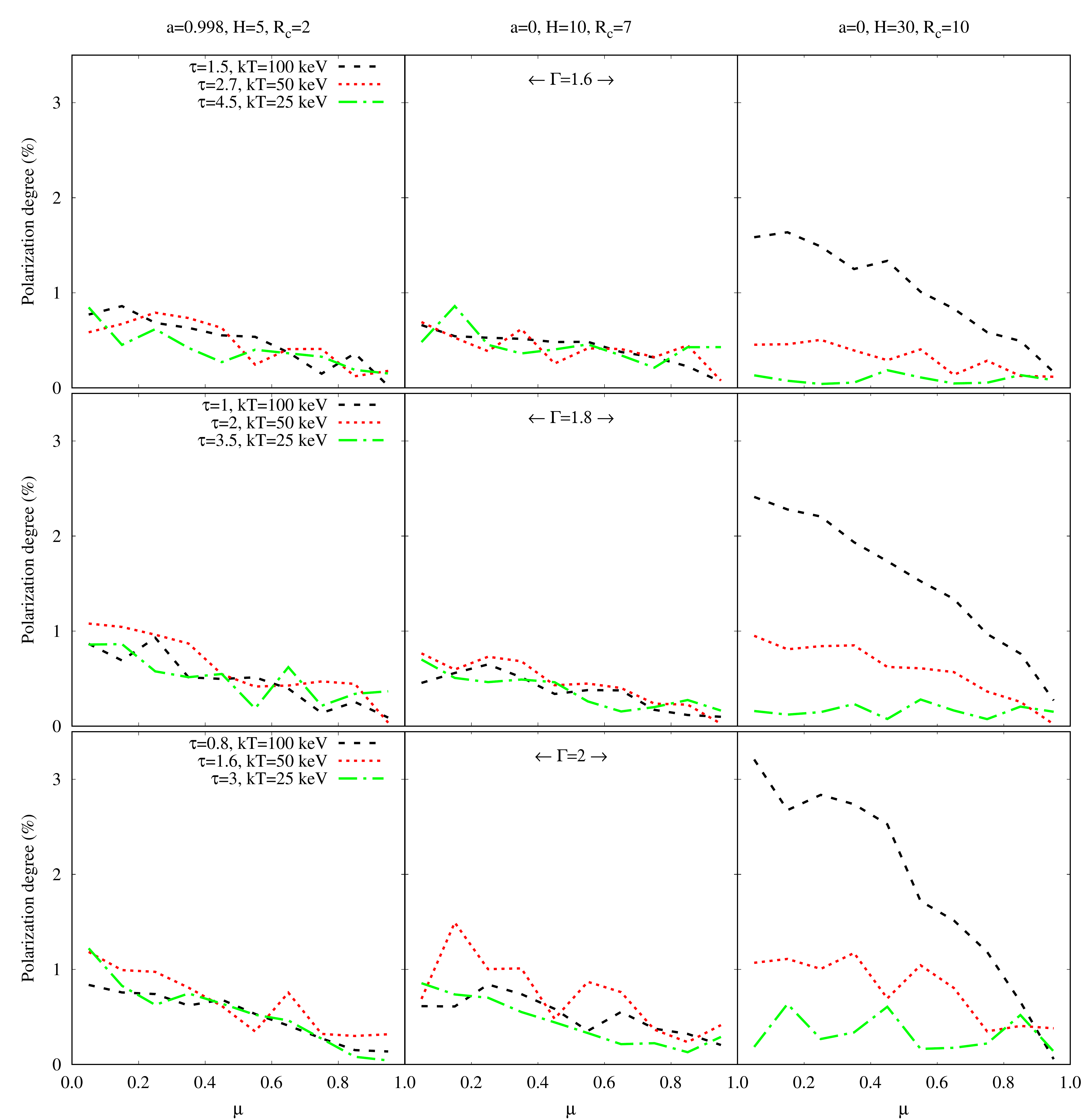}
	\caption{Polarization degree versus the cosine of the inclination angle, for the spherical geometry. Different columns correspond to different values of the height and radius of the corona. Note the different $y$-axis scale compared with Fig. \ref{fig:deg-slab}. \label{fig:deg-sphere}}
\end{figure*}

\begin{figure*}
	\centering Spherical lamppost geometry \par\smallskip
	\includegraphics[width=\textwidth]{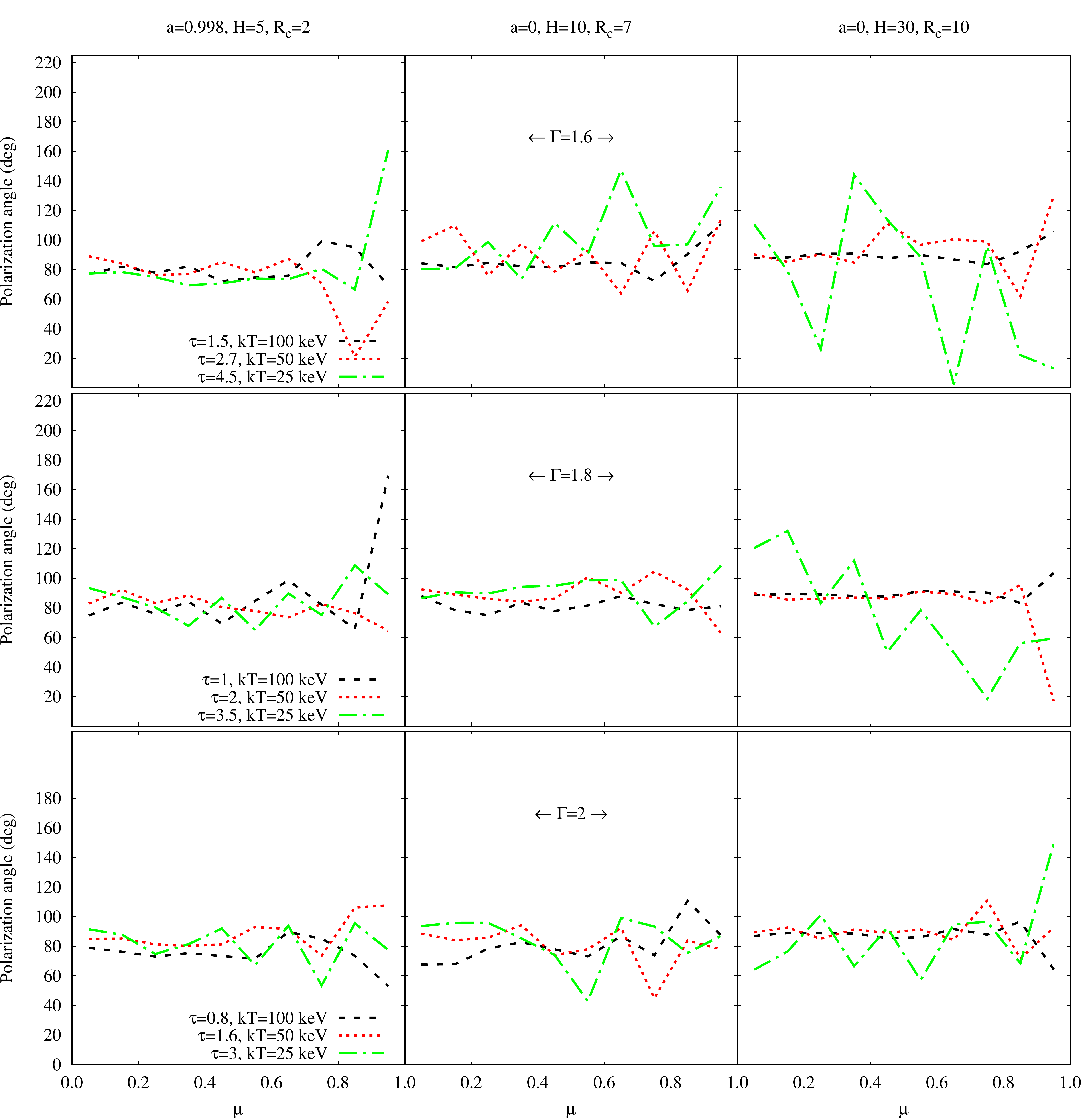}
	\caption{Polarization angle versus the cosine of the inclination angle, for the spherical geometry. \label{fig:ang-sphere}}
\end{figure*}

\begin{figure*}
	\centering Conical geometry \par\smallskip
	\includegraphics[width=\textwidth]{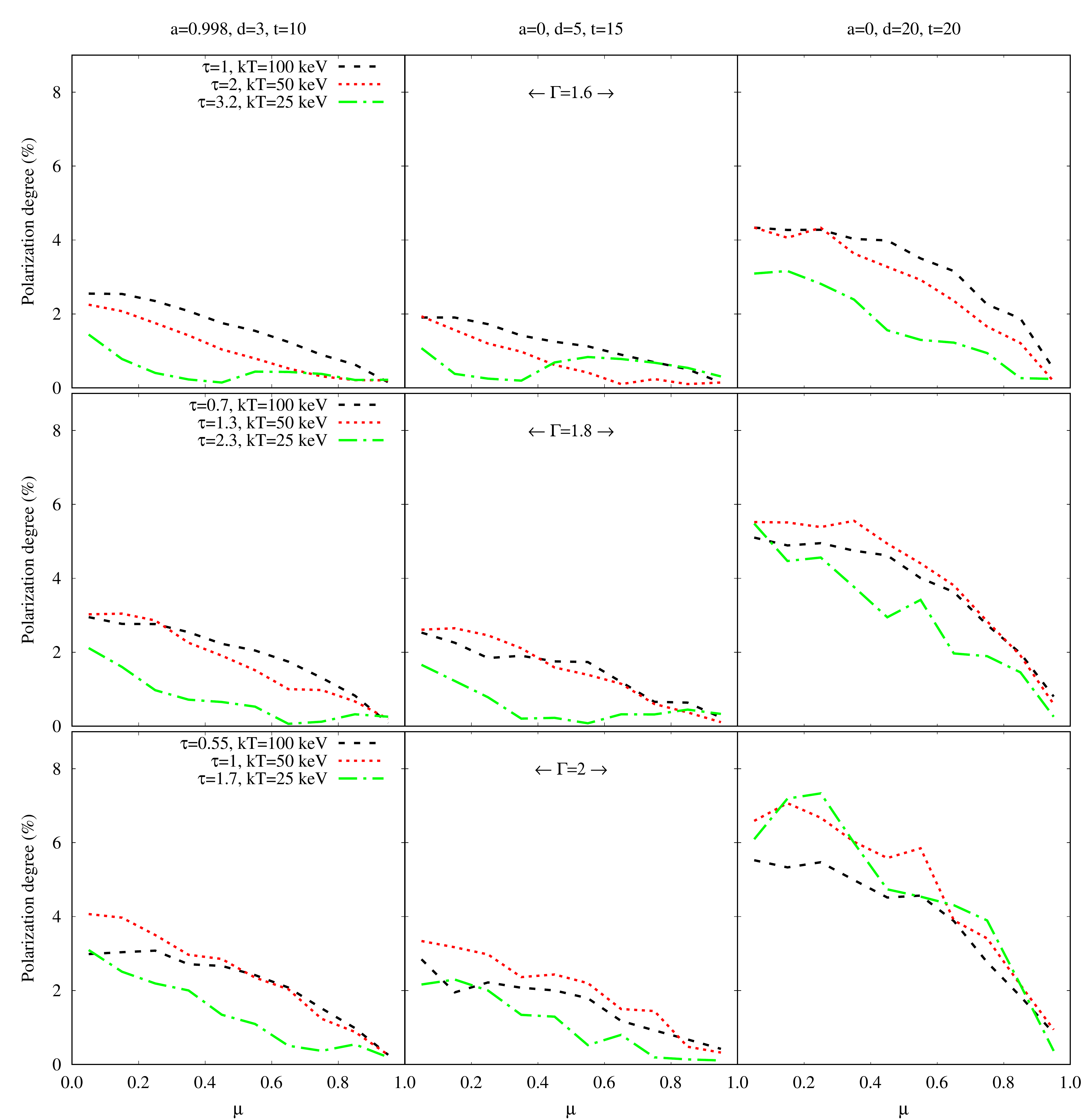}
	\caption{Polarization degree versus the cosine of the inclination angle, for the conical geometry. \label{fig:deg-cone}}
\end{figure*}

\begin{figure*}
	\centering Conical geometry \par\smallskip
	\includegraphics[width=\textwidth]{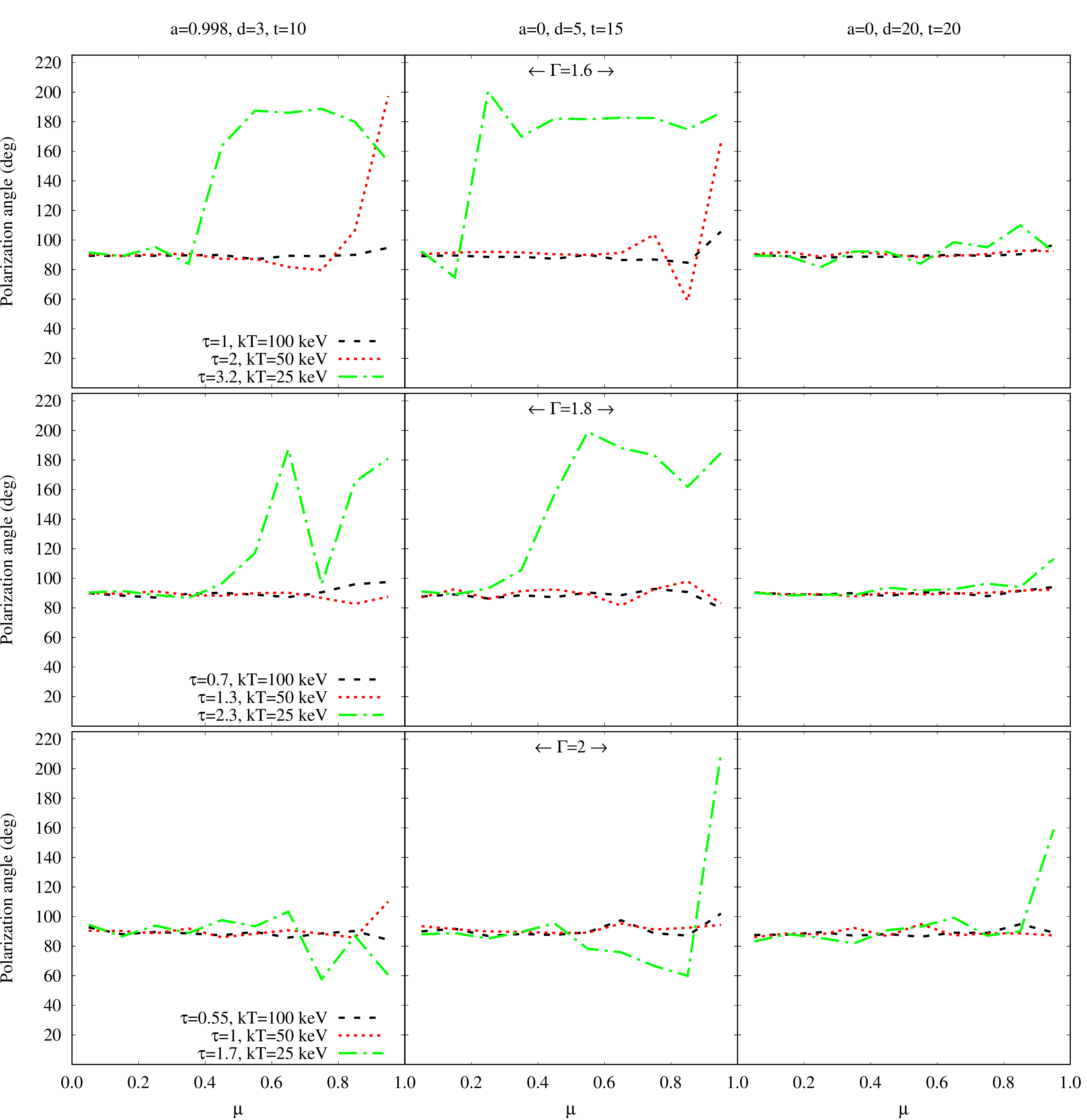}
	\caption{Polarization angle versus the cosine of the inclination angle, for the conical geometry. \label{fig:ang-cone}}
\end{figure*}

In Figs. \ref{fig:deg-slab}, \ref{fig:ang-slab}, \ref{fig:deg-sphere}, \ref{fig:ang-sphere}, \ref{fig:deg-cone}, and \ref{fig:ang-cone}, we plot the polarization degree and angle, as a function of the inclination, for the three geometries considered in this work and for the different coronal parameters. We separate the three geometries to avoid overcrowding the plots. In these figures, different columns correspond to different coronal sizes. Interestingly, the different sizes yield different polarization degrees, as shown by
Fig. \ref{fig:deg-slab} for the slab, Fig. \ref{fig:deg-sphere} for the sphere, and Fig. \ref{fig:deg-cone} for the cone. Concerning the polarization angle, Fig. \ref{fig:ang-sphere} shows some scatter around 90 deg in the spherical case. As argued in Sect. \ref{subsec:angle}, this is not surprising given the low polarization degree. In the conical case, the polarization angle sometimes flips from 90 to 180 deg, especially when the optical depth is large and for observers at low inclinations (see Fig. \ref{fig:ang-cone}). This could be due to the increasing contribution by multiply scattered photons \cite[e.g.][]{tamborra2018}, combined with the decreasing horizontal polarization of the disc photons for observers at low inclinations \cite[e.g.][]{SK2010}. 

\bsp	
\label{lastpage}
\end{document}